\documentclass{elsart}
\usepackage{epsf}
\def\beq{\begin{equation}}
\def\eeq{\end{equation}}
\def\bea{\begin{eqnarray}}
\def\eea{\end{eqnarray}}
\def\max{{\rm max }}

\def\geom{{\rm geom }}
\def\gev{{\rm GeV }}
\def\tev{{\rm TeV }}
\def\susy{{\rm SUSY }}

\def\res{{\rm res }}
\def\to{\rightarrow}

\def\simgt{\,{\rlap{\lower 3.5pt\hbox{$\mathchar\sim$}}\raise 1pt\hbox{$>$}}\,}
\def\simlt{\,{\rlap{\lower 3.5pt\hbox{$\mathchar\sim$}}\raise 1pt\hbox{$<$}}\,}
\def\aa{\ifmmode{ \gamma\gamma } \else{ $\gamma\gamma$ } \fi}
\def\ea{\ifmmode{ e\gamma } \else{ $e\gamma$ } \fi}
\def\epem{\ifmmode{ e^+ e^- } \else{ $e^+ e^-$ } \fi}
\def\emem{\ifmmode{ e^- e^- } \else{ $e^- e^-$ } \fi}
\def\ttbar{\ifmmode{ t\bar{t} } \else{ $t\bar{t}$ } \fi}
\def\qqbar{\ifmmode{ q\bar{q} } \else{ $q\bar{q}$ } \fi}
\def\ffbar{\ifmmode{ f\bar{f} } \else{ $f\bar{f}$ } \fi}
\def\se{\ifmmode{ \tilde{e} } \else{ $\tilde{e}$ } \fi}
\def\snu{\ifmmode{ \tilde{\nu} } \else{ $\tilde{\nu}$ } \fi}
\begin{document}
%\runauthor{Kaoru Hagiwara}
\vspace*{-10mm}
\begin{flushright}
KEK-TH-731 \\
10 November 2000 \\[5mm]
\end{flushright}
\begin{frontmatter}
\title{
Higgs, SUSY and the Standard Model \\ 
at $\gamma\gamma$ Colliders\thanksref{talk}}
\thanks[talk]{Invited talk presented at the \aa 2000 Workshop, 
14--17 Jun 2000, DESY, Hamburg.}
\author{Kaoru Hagiwara}
\address{KEK Theory Group, Tsukuba, Ibaraki 305-0801, Japan}
\begin{abstract}
In this report I surveyed physics potential of the \aa 
option of a Linear \epem Collider with the following questions 
in mind: 
{\it What new discovery can be expected at a \aa collider in addition 
to what will be learned at its `parent' \epem Linear Collider? }
By taking account of the hard energy spectrum and polarization of colliding 
photons, produced by Compton back-scattering of laser light off 
incoming $e^-$ beams, we find 
that a \aa collider is most powerful when new physics 
appears in the neutral spin-zero channel at an invariant mass below 
about 80\% of the c.m.\ energy of the colliding \emem 
system.  If a light Higgs boson exists, its properties can be studied 
in detail, and if its heavier partners or a heavy Higgs boson exists 
in the above mass range, they may be discovered at a \aa 
collider.  CP property of the scalar sector can be explored in 
detail by making use of linear polarization of the colliding photons, 
decay angular correlations of final state particles, and the pattern  
of interference with the Standard model amplitudes.  
A few comments are given for SUSY particle studies at a \aa 
collider, where a pair of charged spinless particles is produced in 
the $s$-wave near the threshold.  Squark-onium may be discovered.  
An $e^\pm \gamma$ collision mode may measure the Higgs-$Z$-$\gamma$ 
coupling accurately and probe flavor oscillations in the slepton sector.  
As a general remark, all the Standard Model background simulation tools 
should be prepared in the helicity amplitude level, so that simulation 
can be performed for an arbitrary set of Stokes parameters of the 
incoming photon beams.  
\end{abstract}
\end{frontmatter}

\section{Why do we need a Photon Linear Collider?}

The photon linear collider (PLC) makes use of the hard energy 
spectrum of the photons produced by Compton backscattering of 
a high power laser light off the linac $e^-$ beam%
\cite{gkst83,gkst84,telnov90}.  
Therefore, a PLC should be considered as an option of a future 
\epem Linear Collider.  
This observation naturally leads us to the following questions:
\begin{itemize}
\item{\it
What new discovery can we expect at a Photon Linear Collider in 
addition to what we will learn at its `parent' \epem Linear 
Collider?
}
\item{\it
Does the PLC option make a Linear Collider project more attractive?
}
\end{itemize}
In this report, I try to find the answer to the above two questions.  

\begin{figure}[p]%[htbp]
\begin{center}
\vspace*{-10mm}
\epsfxsize=135mm
\epsfysize=180mm
\epsffile{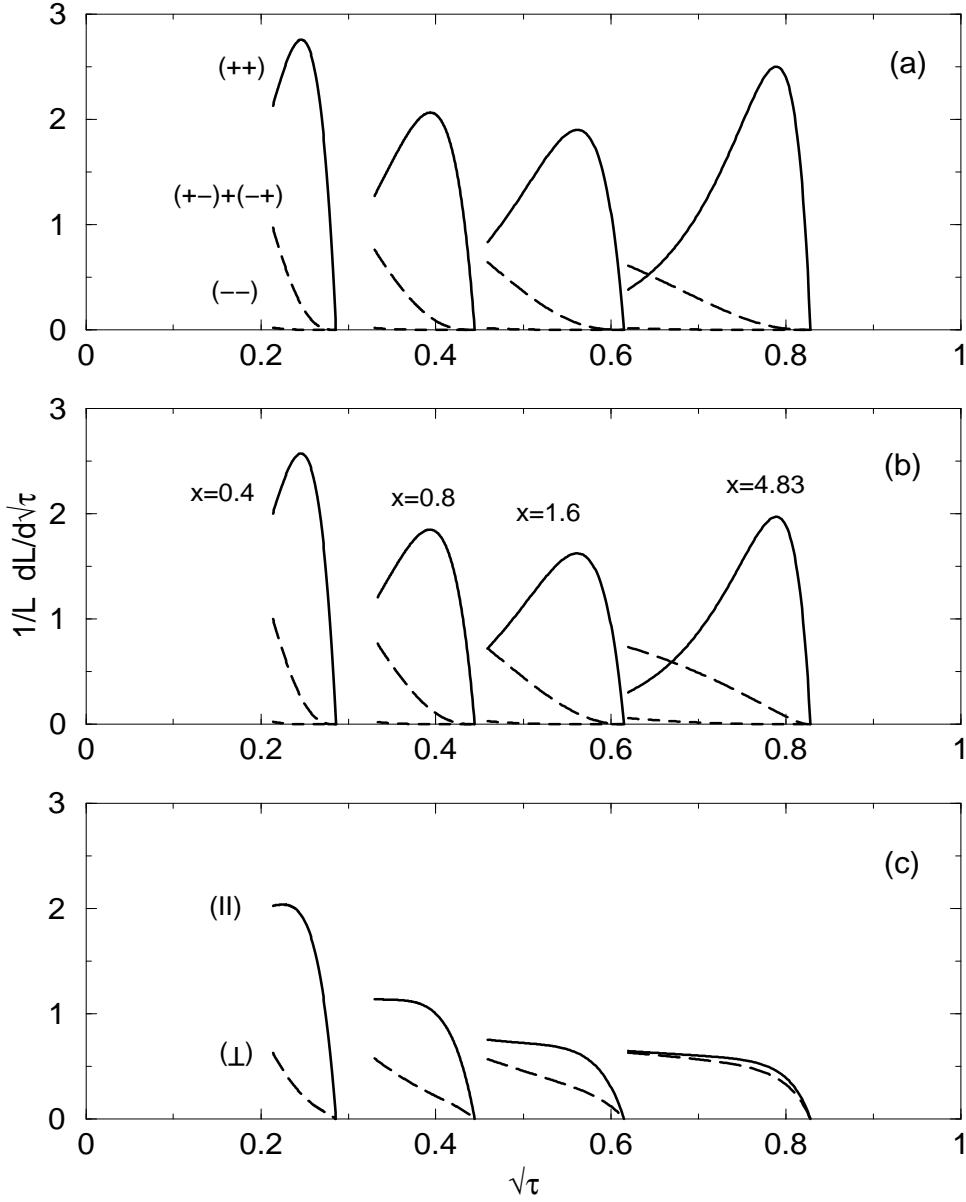}
\end{center}
\caption{{\small 
Normalized luminosity functions of colliding photons as a function 
of $\sqrt{\tau}=\sqrt{s_{\gamma\gamma}}/\sqrt{s_{ee}}$, calculated in the  
Compton back-scattering limit.  Four values of laser frequencies ($w_o$) 
are chosen, $x=4E_e w_o /m_e^2=4.83$, 1.6, 0.8 and 0.4 from the right 
to the left.  In (a) and (b), the luminosity is shown separately for 
the collisions of two right-handed photons denoted by solid lines 
$(++)$, those of right- and left-handed photons by long-dashed lines 
$(+-)+(-+)$, and those of two left-handed photons by short-dashed 
lines $(--)$.  In (a), both photons are obtained by setting 
$P_e P_c=-1$, whereas in (b), they are obtained for 
$P_eP_c=-|P_e|=-0.8$.   
The curves in (c) are obtained for $P_e=0$ and $P_t=1$, when the 
two laser lights have parallel linear polarizations.    
Solid lines show collisions 
when two photons have parallel linear polarizations, and dashed 
lines are for two photons with perpendicular linear polarizations.
}}
\label{fig:aalumi}
\end{figure}

I start my study by comparing the $\gamma\gamma$ channel with the 
$e^+e^-$ annihilation channel, which are both source of various 
new particles.  All charged particles are pair produced in both 
channels, and neutral particles can be produced as $s$-channel 
resonances.  There is, however, an important difference in the 
spin of the accessible $s$-channel resonances.  The $e^+e^-$ 
annihilation channel cannot couple to a spin 0 resonance 
because of the electronic chirality conservation, whose 
breaking is suppressed by the tiny electron mass.  The lowest 
spin of a particle that can be produced in the $e^+e^-$ 
annihilation channel should hence be 1. 
On the other hand, the $\gamma\gamma$ channel can couple to 
a spin 0 resonance, while it cannot couple to a spin 1 resonance 
due to spin statistics of the $J_z=0$ two-photon system\cite{landauyang}.   
It is this stunning difference between the two channels that 
makes a PLC complementary to $e^+e^-$ colliders.  
We can probe the scalar sector in the $s$-channel of the colliding 
two photons at a PLC, whereas it can be probed only in association 
with another particle production at $e^+e^-$ colliders.  
This possibility gives a PLC the unique potential of becoming the 
best observatory of the scalar sector, or the Higgs sector.  
We should take this opportunity very seriously, because the 
scalar sector is the least known sector of the Standard Model (SM), 
and because its detailed understanding is probably the most 
important key in our search for physics beyond the SM.  

There are many excellent reports\cite{watanabe97,watanabe99,jikia99,melles99} 
on the role of a PLC as the laboratory of the Higgs sector, and I will 
give only a few general remarks in section 2.  
In addition to the precision measurements of the two-photon decay partial 
width and branching fractions of the Higgs bosons, I would like to 
emphasize the importance of probing new interactions, including CP 
violating ones, in the scalar sector.  
PLC is particularly well tailored for studying the CP property of 
resonances and interactions, because the two $J_z=0$ two-photon 
initial states can form a CP-even and a CP-odd state, and they can 
be prepared by using linear polarizations of the laser beams.
The power and limitation of PLC with linearly polarized laser beams 
are discussed.  
Once the Higgs property is determined well in $\gamma\gamma$ collisions, 
measurement of the Higgs-$\gamma$-$Z$ coupling may be done in the 
$e\gamma$ collision mode\cite{chois95}.  

The role of a PLC in our study of supersymmetry is discussed in 
section 3. 
Because the available highest c.m.\ energy in the $\gamma\gamma$ 
mode is about 80\% of the original $e^+e^-$ collision c.m.\ energy, 
\beq
\sqrt{\tau} = \frac{\sqrt{s_\aa }}{\sqrt{s_{ee}}} < \frac{x}{x+1} 
< \frac{2+2\sqrt{2}}{3+2\sqrt{2}} \approx 0.83\,,
\label{def:tau}
\eeq
one generally expects that the $\epem$ mode is the discovery channel 
for those SUSY particles which may escape detection at the LHC.  
Here $x=4E_e w_o /m_e^2$ is the normalized laser frequency  
which is bounded from above, $x<2+2\sqrt{2}$, in order not to loose 
the photon beam by soft $\epem$ pair production.  
A PLC should therefore provide us with measurements which cannot 
be matched by the other experiments.  
Formation of squark-onia, precision measurements of SUSY particle  
properties, 
and measurements of CP violation in the chargino sector are examined.  
In the $\ea$ mode, single production of s-electron in association 
with a neutralino may turn out to be the best laboratory of 
s-lepton flavor physics. 
The \emem mode can be the precision SUSY factory if s-electron 
pair can be produced.  

In section 4, I give a few remarks on the role of a PLC in the study 
of the properties of the SM particles and their interactions.  
Such studies could be the main theme of high energy physics if 
neither the LHC nor $\epem$ collider fail to identify the physics 
beyond the SM.  
I also emphasize the importance of preparing all the SM background 
simulation programs in the helicity amplitude level so that they can 
be simulated for arbitrary polarization of the colliding photons. 

Throughout this report, I use a very simple approximation to a PLC 
where the Compton scattering formula is used to generate colliding 
photons in the exactly backward direction.   
This description gives a good approximation only for the hard part 
of the photon beams\cite{ginzburg00,cain}.  
In Fig.~\ref{fig:aalumi}, I show the normalized effective luminosity 
function 
\beq
\frac{1}{L_\aa }\frac{d L_\aa }{d \sqrt{\tau}}
\label{daalumi}
\eeq
in this approximation for four values of the laser frequency parameter 
$x$, $x=2+2\sqrt{2}$, 1.6, 0.8 and 0.4 from the right to the left, 
and for three cases of the electron and laser polarizations, (a) to (c).  
Only the region where $\sqrt{\tau}/\sqrt{\tau}_\max > 0.75$
are shown, where the approximation may hold.  
In (a) and (b), circularly polarized laser beams ($P_c=\pm 1$) 
are used to make collisions of definite helicity photons, 
and the luminosity distribution is given 
for the collision of two right-handed photons $(++)$, denoted by solid 
lines, that of right- and left-handed photons $(+-)+(-+)$, denoted by 
long-dashed lines, and that of two left-handed photons $(--)$, 
denoted by short-dashed lines.  
In (a), both photons are obtained by setting $P_e P_c=-1$, whereas 
in (b) they are obtained by setting $P_e P_c=-P_e=-0.8$, as a more 
realistic case.  
Although the laser light may be 100\% circularly polarized ($|P_c|=1$), 
the colliding \emem beams have finite polarization.  
It is worth noting here that nearly optimal monochromaticity of the 
colliding photon polarizations is obtained with the electron beam 
polarization of $|P_e|=0.8$ which may well be realized.  
The curves in (c) are obtained for $P_e=0$ and $P_t=1$  
when the two laser photons have parallel linear polarization. 
Here $P_t$ stands for the degree of linear polarization.  
Solid lines show collisions when two photons have parallel linear 
polarizations, and dashed lines are for perpendicular linear 
polarizations. 
The former two-photon state is CP-even while the latter state 
is CP-odd.  
We can clearly see that capability of distinguishing the two cases 
is small for high laser frequencies ($x$) or when 
$\sqrt{\tau}\simgt 0.6$.

Before closing this section, I should note that the actual luminosity 
function may depend strongly on the property of the incoming electron 
beams and on the electron-to-photon conversion efficiency.  
We may express 
\bea
\frac{dL_\aa }{d\sqrt{\tau}} &=& \kappa^2 L_{ee}^\geom 
\bigl( \frac{1}{L_\aa }\frac{d L_\aa }{d \sqrt{\tau}} \bigr)\,, 
\label{aalumi}\\
\frac{dL_\ea }{d\sqrt{\tau}} &=& \kappa L_{ee}^\geom 
\bigl( \frac{1}{L_\ea }\frac{d L_\ea }{d \sqrt{\tau}} \bigr)\,, 
\label{ealumi}
\eea
where $\sqrt{\tau}=\sqrt{s_\aa }/\sqrt{s_{ee}}$ in Eq.~(\ref{aalumi}) 
while $\sqrt{\tau}=\sqrt{s_\ea }/\sqrt{s_{ee}}$ in Eq.~(\ref{ealumi}).  
$L_{ee}^\geom$ is the geometric luminosity of the colliding 
\emem beams in the absence of beam-beam effects, 
and $\kappa$ denotes the conversion factor of order 0.5.  
Because the geometric luminosity can be significantly larger than 
actual collision luminosity in the \epem or \emem modes, 
there is a possibility that the luminosity integrated over the 
high $\sqrt{\tau}/\sqrt{\tau}_\max $ region shown in 
Fig.~\ref{fig:aalumi} can be comparable or even larger than the 
corresponding luminosity of \epem collisions.  
When we compare physics capability of a PLC with its parent \epem 
or \emem LC, I assume that the integrated $\aa$ luminosity in the high 
$\sqrt{\tau}/\sqrt{\tau}_\max $ region is about 
the same as the luminosity of \epem collisions.  

It should be remarked here that the luminosity distributions as shown 
in Fig.~1 do not include contributions from the beamstrahlung.  
At a PLC, since the electron beams are tuned to maximize the 
\aa luminosity, more beamstrahlung may be produced than in the 
corresponding \epem collisions.  
It is therefore important to estimate the effects of beamstrahlung 
in all quantitative studies, especially in the relatively low 
$\sqrt{\tau}$ region where we expect to have high degree of linear 
polarization transfer.  

\section{Neutral Higgs bosons} 

A PLC associated with a 500~GeV \epem LC will have the strongest 
case when a light Higgs boson exists below 200~GeV.  
Despite failures of discovering the Higgs boson so far, this 
is the most favored scenario of particle physics at the moment, 
because it is favored by the electroweak precision measurements%
\cite{lepewwg00}, and because it is predicted by all supersymmetric 
theories with grand unification of the three gauge couplings%
\cite{susymh}.  
Such a Higgs boson may still be found at LEP200 or at Tevatron 
if its mass is very near to the present lower mass bound ($\sim 110\gev$) 
and if it has nearly maximum coupling to the $W$ and $Z$ bosons.  
It should certainly be discovered at LHC through gluon-gluon or 
$WW/ZZ$ fusion, unless its couplings to gluons and $WW/ZZ$ are 
both very small, or if it has little decay branching fractions 
into all the observable channels, such as \aa, $\tau^+\tau^-$ and 
$WW^*/ZZ^*$.  
It should be definitely discovered at an \epem LC as long as it 
has a significant coupling to the $W$ and $Z$ bosons, the condition 
which is required for a good fit to the electroweak data\footnote{
Otherwise, there should be some sort of cancellation between the 
heavy Higgs boson and new physics contributions to the precisely 
measured electroweak parameters.}   
and for the perturbative unification of the gauge couplings 
in SUSY models.   

A 500~GeV LC will measure its couplings to $Z$ and $W$ bosons very 
accurately, and measure its decay branching fractions\cite{higgsatlc}, 
Br$(bb)$, Br$(\tau^+\tau^-)$ and Br$(WW^*)$ with a good accuracy, 
as well as Br$(cc)$ and Br$(gg)$.  
The PLC will in addition measure the partial width $\Gamma(H\to\aa)$ 
at a few \% level\cite{jikia99,melles99}.  
Because the $H\aa$ coupling receives contributions from all the 
massive charged particles that couple to the Higgs boson, its 
accurate measurement will give us decisive information on 
the scalar sector when combined with accurate measurements in  
\epem collisions.  
For instance, by using the couplings that are measured in \epem 
collisions, one may estimate the Higgs coupling to the top-quarks and 
other new states by using the $\Gamma(H\to\aa)$ data. 

I note in passing that the Higgs-$Z\gamma$ coupling can be most 
accurately measured in the $e\gamma$ collision mode of a 
PLC\cite{chois95}.  
It is clear that accurate knowledge of both the Higgs-$\aa$ and 
the Higgs-$Z\gamma$ couplings will be powerful in probing the 
quantum numbers of the charged particles whose masses originate 
from the electroweak symmetry breakdown. 

In addition, I would like to emphasize the importance of the 
CP measurements in the scalar sector.  
Because the gauge interactions do not allow CP violation,\footnote{
Disregarding the CP-odd vacuum angle of QCD, whose effect is 
known to be negligibly small.}  
non-gauge interactions should be responsible for the observed 
CP violation, and ultimately, for the origin of the matter 
dominated universe.  
The non-gauge interactions among scalar bosons and between 
scalars and fermions are among the most likely sources of CP 
violation, and precise measurements of the CP properties of 
the Higgs bosons and their interactions may open a completely 
new road in our investigation.  
The PLC will be an excellent laboratory for CP violation in 
the scalar sector, because it allows us to prepare the $J_z=0$ 
initial states with definite CP parity.  
By denoting the two colliding photon helicities as $\lambda_1$ 
and $\lambda_2$, the CP transformation changes their signs 
\beq
\bf{CP} \left\vert\,  \lambda_1,\,  \lambda_2 \right> = 
         \left\vert\, -\lambda_1,\, -\lambda_2 \right> 
\label{cpaa}
\eeq
and hence the two states with definite CP parity are obtained as 
\beq
\bf{CP} \big( \left\vert\, ++ \right> \pm \left\vert\, -- \right> \big)  
= \pm\,  \big( \left\vert\, ++ \right> \pm \left\vert\, -- \right> \big)\,.
\label{cpaa2}
\eeq
The CP-even state has two linearly polarized photons with the polarization 
planes parallel, while the CP-odd state has perpendicular linear 
polarization directions.  
If CP is a good symmetry of the scalar sector, the CP-even Higgs boson 
can couple only to the CP-even initial state, whereas the CP-odd Higgs 
boson can couple only to the CP-odd initial state.   

In Fig.~1(c), I show the \aa luminosity when the two laser beams are 
both linearly polarized ($P_t=1$) along the same direction in the perfect 
backward scattering configuration.  
The Compton scattering with unpolarized \emem beams then produces 
collisions of high energy linearly polarized photons which are partially 
CP-even (parallel, or $\parallel$) and partially CP-odd (perpendicular, 
or $\perp$).  
When the initial laser polarization planes are made to the perpendicular 
orientation, the luminosity functions of the CP-even and CP-odd 
configurations are reversed.  
We note that in this simple Compton scattering scheme, the difference 
between CP-even and CP-odd luminosity functions is significant only 
for relatively low laser frequencies ($x\simlt 1.6$) or at relatively 
low \aa invariant mass, $\sqrt{\tau}\simlt 0.6$.  
The linear polarization of the PLC will hence be useful for studying 
the CP property of the neutral scalar boson whose mass is less than 
about 50\% of the initial \emem collision energy, $\sqrt{s}_{ee}$.    
Because of the necessity of relatively low $\sqrt{\tau}$ values to 
achieve high degree of linear polarization transfer, backgrounds 
from beamstrahlung photons should be estimated in quantitative 
studies.

It is worth noting here that it is an easy task for \epem LC to 
distinguish between a CP-even and CP-odd neutral Higgs bosons. 
Such discrimination can e.g. be made by a simple angular correlation 
study in the process $\epem \to ZH$ followed by the decays 
$Z\to \ffbar$\cite{hs93,kksz94,hikk00}.  
What is difficult in the \epem mode is to detect small CP violation 
effects in the study of dominantly CP-even Higgs bosons.  
The observable effects are expected to be rather small in \epem collisions 
because the small CP-odd component can contribute to the process only 
in the one-loop order whereas the dominant CP-even component 
contributes at the tree-level.  
At a PLC, both components are expected to contribute in the one-loop 
order, and hence we can generally expect bigger CP asymmetries\cite{gunion94}.  
The use of the linearly polarized laser light allows us to make 
the precision CP measurement a counting experiment when the Higgs boson 
mass is less than about 50\% of the \epem collision energy\cite{choilee99}.  
Although we may need to rely more on the final state decay angular 
correlation studies for higher mass bosons, the advantage of larger 
CP asymmetry expected at a PLC will persist for all the neutral spin-less 
states that couple to the \aa channel.  

A PLC will be powerful in studying/discovering the neutral Higgs 
bosons (or its partners) which have suppressed couplings to the 
$Z$ and $W$ bosons\cite{aksw99}.  
Such states are expected in multiple Higgs doublet models including 
the SUSY-SM, and in fact their existence at or below the TeV scale  
makes the unification of the three gauge couplings possible in the 
minimal SUSY-SM\cite{mssmunif}.  
The precision electroweak experiments constrain the mass of the Higgs 
boson which has significant couplings to the $W$ and $Z$ bosons to be 
less than about 200~GeV, or else there should be subtle cancellation 
among new physics contributions.  The degree of subtleness of this 
cancellation increases as the mass of the Higgs boson increases.    
Therefore it is most natural for us to expect that a light Higgs boson 
of mass in the range $100\sim 200~\gev$ has nearly the maximal couplings 
to the $W$ and $Z$ bosons, and hence its heavier partners do not have 
significant couplings to the weak bosons.  
Such states are difficult to produce singly at \epem collisions, and 
they can be discovered at a PLC if their masses lie in the range 
$0.5\sqrt{s_{ee}}\simlt m_{\rm Higgs} \simlt 0.8\sqrt{s_{ee}}$.  

As an example of how heavier Higgs bosons may be found at a PLC, 
I show in Fig.~2(a) the cross section of the process 
\beq
\aa \to \ttbar
\label{aatt}
\eeq
as a function of the invariant mass of the final state, $m(\ttbar)$.  
The cross section is calculated for a 500~GeV \epem LC with the \aa 
luminosity function of Fig.~1(a) at the highest laser frequency 
($x=4.83$).  
It should be noted that because of the high \ttbar threshold 
($2m_t/\sqrt{s_{ee}}\approx 0.7$), only the peak region of the 
\aa luminosity functions contributes where the purity  
of the collisions of right-handed photons ($++$) is high.   
The thick short dashed line shows the prediction of QED.  
The thick solid line (long-dashed line) shows the prediction 
where a CP-odd (CP-even) scalar boson of mass 400~GeV 
is produced as an $s$-channel resonance.  
For definiteness, we use the MSSM (minimal SUSY-SM) prediction for 
the total and partial widths of a 400~GeV CP-odd Higgs boson, 
A\footnote{% 
$\Gamma_A=1.75~\gev$, Br$(A\to\ttbar)=0.95$, 
Br$(A\to\aa)=1.5\times 10^{-5}$ for 
$\tan\beta=3$, $m_\susy=1~\tev$, $M_2=500~\gev$ and $\mu=-500~\gev$  
as chosen in ref.~\cite{aksw99}.}.      
As a comparison, predictions for the CP-even Higgs boson case 
are obtained simply by reversing the CP-parity of the state while 
keeping all the other properties.  
It is clearly seen that the interference pattern with the QED 
amplitude is very sensitive to the CP-parity of the resonance.  

\begin{figure}[p]
\begin{center}
\vspace*{-10mm}
\epsfxsize=135mm
\epsfysize=180mm
\epsffile{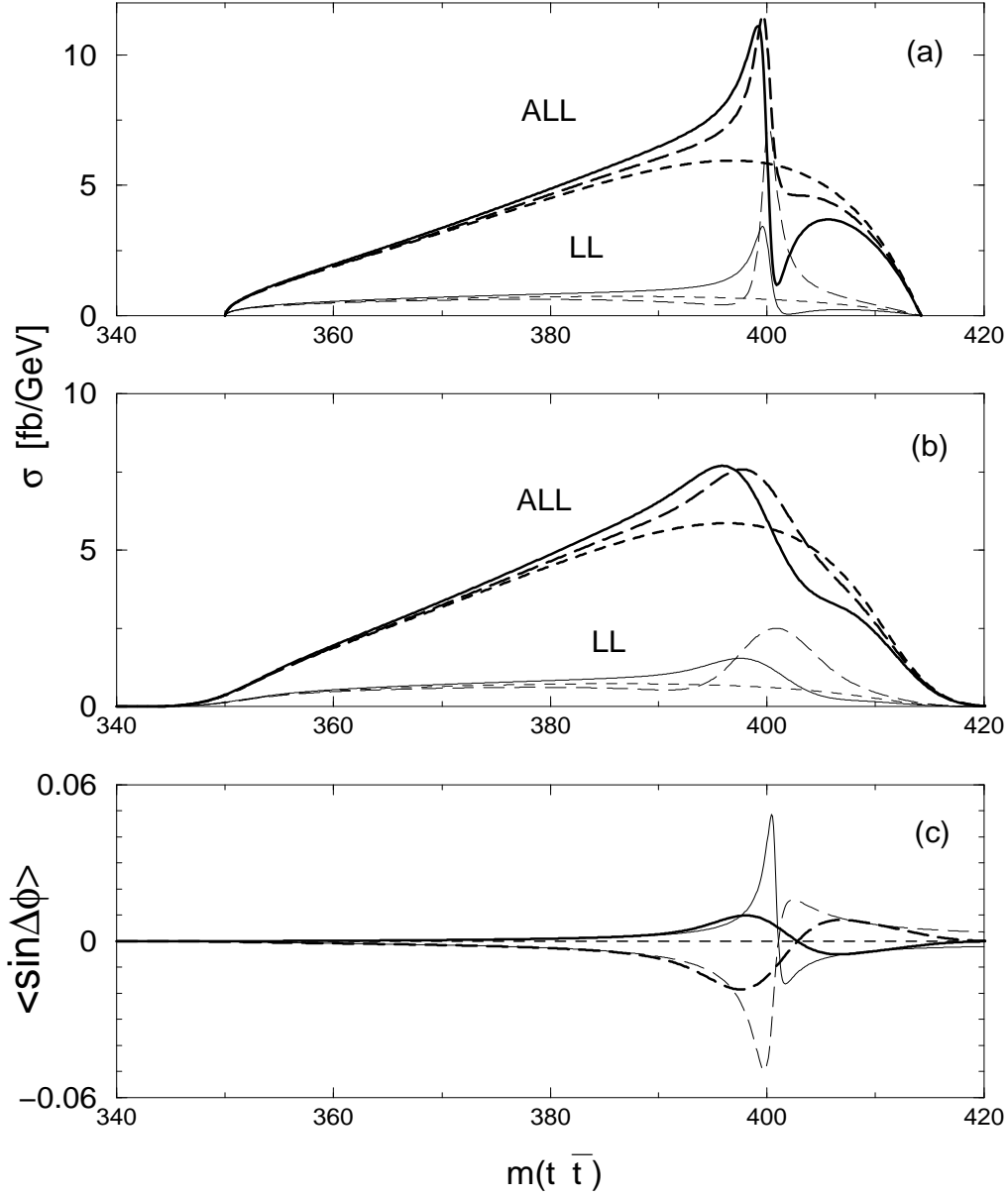}
\end{center}
\caption{{\small 
The cross section of $\aa \to \ttbar$ events as a function of 
$m(\ttbar)$ for the $\aa$ luminosity of Fig.~1(a) for $x=4.83$ 
at $\sqrt{s}_{ee}=500~\gev$.  
Short-dashed curves show QED predictions, 
while solid (long-dashed) curves show predictions when a 400~GeV 
CP-odd (CP-even) spin-less resonance is produced in the $s$-channel.  
In (a) and (b), the thick lines are for total events and the thin 
lines are for events where both $t$ and $\bar{t}$ are left-handed.  
Gaussian smearing with $\Delta m(\ttbar)=3~\gev$ is applied in (b). 
Azimuthal decay angular correlation is shown in (c) with (without) 
the smearing by thick (thin) lines.     
}}
\label{fig:aatt}
\end{figure}

In fact the whole difference appears in the helicity amplitudes 
$\gamma_+ \gamma_+ \to t_L \bar{t}_L$ where both $t$ and $\bar{t}$ 
quarks are left-handed. 
The thick lines show the distributions when all the \ttbar helicities 
are summed up, whereas the thin lines show those when the 
$t_L \bar{t}_L$ events (denoted as `LL') are selected.  
Note, however, that even though the signal is clearer when 
the LL events are selected, only those events when one of the $W$'s 
decay leptonically can be used to distinguish the LL events from 
the dominant RR events.  
Still, about 40\% of all the \ttbar events may be used for such 
helicity analysis.  
It may be worth noting here that even the 6-jet events can be used 
to distinguish between the LL+RR modes and the LR+RL modes.  
In our example, because of suppressed $J_z=\pm 2$ \aa luminosity 
distribution in Fig.~1(a), very small fraction of all the \ttbar 
pairs have the polarization LR+RL.  

A more serious problem at a PLC is that the invariant mass of the 
colliding \aa system, $\sqrt{s_\aa }=m(\ttbar )$, can only be 
measured through the final top-quark momentum measurement.    
Accurate knowledge of the top-quark mass and their decay properties 
which will be obtained in the \epem collision experiments will be 
used to refine such measurements.  
Fig.~2(b) shows the $m(\ttbar )$ distributions when a Gaussian 
smearing with the error
\beq
\Delta m(\ttbar ) = 3~\gev 
\label{mtterror}
\eeq
is applied.  
Comparisons between the distributions of Fig.~2(a) and Fig.~2(b) show 
that the sharp peaks and dips in the original curves are smeared out and 
accurate knowledge of the measurement error, $\Delta m(\ttbar )$, 
will be needed to determine 
the mass and the widths of the Higgs resonances.  
The quoted error of 3~GeV may be too optimistic especially for 
semi-leptonic modes where at least one hard neutrino is missing.  
More work is needed to make the error as small as possible, 
for both 6-jet and semi-leptonic modes.   

From Fig.~1(c), we find that linearly polarized laser beams produce 
little CP discriminating power at large $x$, or at large $\sqrt{\tau}$.  
When a scalar resonance is found in the mass range of 
$0.6 \simlt m_\res /\sqrt{s}_{ee} \simlt 0.83$, we should refer to 
the final state analysis to determine its CP property.  
As an example of CP-sensitive observables, I show in Fig.~2(c) the 
$m(\ttbar )$ dependences of $\left< \sin\Delta\phi \right>$, 
where $\Delta\phi$ stands for the difference of the azimuthal 
angles of the $t$-decay and $\bar{t}$-decay planes along the 
\ttbar momentum axis.  
The thin lines are predictions before smearing and the thick 
lines after smearing with the error of Eq.~(\ref{mtterror}).  
The asymmetry has opposite signs between the CP-odd and 
the CP-even resonances, whose predictions are shown by the solid 
and the long-dashed curves, respectively.  
The asymmetry is found to be rather small because of cancellation 
between the contributions from longitudinally and transversally 
polarized $W$'s.  
One can certainly find final-state observables which have 
significantly higher sensitivity to the CP-properties of the 
resonance, by taking account of the $W$-decay distributions.  

In summary, if the resonance mass is low enough, say 
$m_\res /\sqrt{s}_{ee} \simlt 0.6$, we can use the linear polarization 
of laser beams to determine its CP property.  
If the resonance has significant decay branching fraction into
spinful heavy particles, such as \ttbar or $W^+W^-/ZZ$, it is 
relatively easy to determine its CP property by making use of 
the decay angular correlations.  
Only when the mass is in the range  
$0.6 \simlt m_\res /\sqrt{s}_{ee} \simlt 0.83$ and when 
it rarely decays into heavy spinful particles, 
we will have difficulty in determining its CP property.  
The possibility of giving high degree of linear polarization to the 
colliding high-energy photons, that has been discussed at 
this workshop\cite{serbo00} may turn out to be useful in such cases.  
I also feel that more serious work may be needed to determine if 
the $\tau^+\tau^-$ decay mode of a heavy resonance can be used 
to study its spin and CP-parity at a PLC.   
In general, it is important to make CP measurements at all possible 
channels, so that we can probe CP-violation in the mixing, 
in the production, and in various decay channels\cite{achl00}.  

\section{SUSY particles and charged Higgs bosons} 

At a PLC, we can produce a pair of squarks, sleptons, charged 
Higgs bosons and charginos above the threshold.  
Because the \aa collision at a PLC can reach the c.m.\ energy 
of at most about 80\% of the corresponding \epem collision energy, 
it is unlikely that these particles are discovered at a PLC.  
The question is what advantage does a PLC have over its parent  
\epem LC when studying their properties.  

In case of squarks, sleptons and charged Higgs bosons, 
we note that the pair can be 
produced in the $s$-wave at a PLC, whereas the pair can only be 
produced in the $p$-wave near threshold at \epem collisions\footnote{
The exception to this rule is the production of 
$\se_L^\pm \se_R^\mp$ pairs where the electronic chilarity 
of the initial channel is transfered to the final state.  
With proper choice of initial $e^\pm$ beam polarizations, 
these pairs can be produced at $s$-wave near the threshold.}.
This can lead to a higher production rate at a PLC depending on 
the factor of $\kappa^2 L_{ee}^\geom$ in Eq.~(\ref{aalumi}). 
It is possible that the production rate larger than that 
at the parent \epem LC can be achieved at a PLC  
if the factor of $\kappa^2 L^\geom_{ee}$ can be made 
significantly larger than the actual \epem luminosity.  
In such cases, precision measurements of the charged scalar 
boson properties may be performed at a PLC.   
In addition, in case of squarks, we may find the $s$-wave $J=0$ 
squark-onia at a PLC.  
There might be a case when the squark-pair production is sensitive to  
the Higgs-boson exchange between stop quarks\cite{hkmn90}.  

In case of chargino-pair production near the threshold, they are 
produced in the $s$-wave both in \epem and \aa collisions.  
The only difference is that the pair is in the spin-triplet state in 
case of \epem collisions while it is in the spin-singlet state at 
a PLC.  
I do not know if this difference leads to a significant difference 
in the study of their properties.  
Because the $s$-wave spin-singlet state is CP-odd, a PLC may be useful 
in probing the CP property of the chargino-photon couplings.  
This however requires a PLC with relatively low laser frequencies 
(hence a higher \epem energy), and the sensitivity should be compared 
with that of the final state analysis in the \epem 
mode\cite{tsukamoto95,choizerwas}.  

There is one point which might be worth noting.  
Both the sfermion-pair and chargino-pair production cross sections 
are uniquely determined by QED in the leading order of \aa collisions.  
This uniqueness of the tree-level amplitudes may help us identify 
radiative effects.  
Certainly a combination of precision measurements of production 
cross sections at both \epem and \aa collisions will give us 
useful additional information on the interactions of the 
SUSY particles and the charged Higgs bosons.  

Finally, the \ea collision mode of a PLC may become a unique 
laboratory of slepton flavor physics\cite{feng96} 
if $\se_L$ or $\se_R$ ($\snu_{eL}$) can be produced in 
association with a neutralino (chargino).  
Flavor oscillation in the slepton sector can be most clearly 
studied in this channel where a SUSY particle state with a 
definite flavor and chilarity quantum numbers can be produced 
at the time of collisions.  

Before concluding this section, let me comment on the possibility 
of precision SUSY tests in the process\cite{feng97} 
$e^-_\alpha e^-_\beta \to \se^-_\alpha \se^-_\beta$ 
for the three distinct channels, 
$\alpha\beta = LL$, $RR$ and $LR$.  
The process is most suited for precision measurements of the 
s-electron masses and their couplings to the gauginos, 
which may give us precious information on the heavy SUSY 
particle masses\cite{hikasa95,nojirim96,feng97}.   
Dedicated study of the \emem collider option is worth serious 
attention, which requires studies independent of the \emem beams 
which are optimized for the PLC option.  

\section{The Standard Model} 

When we consider the SM processes as a probe of new physics, 
I think that a PLC can have an advantage over its parent \epem 
LC when we study in detail the $J=0$ channel, by using 
the monochromaticity of high $\sqrt{\tau}$ region with high 
laser frequency, or when we study the CP property by using the 
linear polarization with relatively low laser frequency.  
Measurements of $W$ and top-quark EDM\cite{ch95} are examples 
of the latter.  
As for the $J=0$ channel, $ZZ$\cite{jikiazz}, $W^+W^-$\cite{jikiaww}  
and $\ttbar$\cite{jikiatt} final states 
are all important because they should couple to the 
electroweak symmetry breaking sector to obtain their masses.  
That the electroweak precision measurements favor the SM 
Higgs boson of mass below about 200~GeV implies that if the Higgs 
boson (whose coupling to $W$ and $Z$ bosons are significant) is 
much heavier than 200~GeV or absent, there should be new physics 
that couple to $W$ and $Z$ bosons.  
Because it should affect the $W$ and $Z$ properties significantly, 
we should be able to identify its effect at LHC or at a lepton 
collider.  
The combined study of the $J=1$ channel at \epem LC and the 
$J=0$ channel at a PLC may be most fruitful in the search of 
such new interactions.  

I would like to note also that detailed studies of purely neutral 
gauge boson scattering processes, 
$\aa \to \aa$, $Z\gamma$ and $ZZ$, may give us useful information 
on new physics that affect these channels either in the tree-level 
or through radiative effects.  
The complete helicity amplitudes for all these mode in the SM 
are known\cite{gounaris99} and they should be useful in determining 
the properties of new physics that affect these processes. 

The SM processes are also important for monitoring of the luminosity 
and the polarization of colliding photons, and also as backgrounds 
in new particle studies.  
Because we will need both circular and linear polarization of 
laser light, it is important for us to prepare all the SM simulation 
tools in the helicity amplitude level.  
All the SM processes should be generated for an arbitrary set of 
the Stokes parameters of the incoming photons.  
For instance, since massless leptons and quarks cannot be produced 
at large scattering angles from $J_z=0$ two photons in the lowest 
order of QED, $J_z=0$ luminosity functions may be monitored by using 
higher order processes, such as $l^+l^-\gamma$, 
$l^\pm \gamma (l^\mp)$ and 
$l^+l^-{l'}^+{l'}^-$, or by using $W^+W^-$\cite{watanabe92}.  
Hadron jet shape from $\aa \to \qqbar (g)$ processes may also be 
sensitive to the ratio of $J_z=0$ and $J_z=\pm2$ collisions 
because we expect $J_z=0$ photons to produce fatter jets.  
The $J_z=0$ luminosity function may be measured more efficiently by 
reversing the laser and electron polarizations simultaneously but  
in one side only, which leaves all the distributions the same while 
replacing the $J_z=0$ and $|J_z|=2$ distributions\cite{telnov93}.  
Linear polarization may be monitored by azimuthal angle distributions 
of high $p_T$ $l^+l^-$ events.  
All these studies should be done in the presence of realistic 
beamstrahlung backgrounds in order to estimate the monitoring errors.  

\section{Conclusions} 

If a light Higgs boson of mass below 200~GeV is found, a PLC should 
be built in association with the first stage of an \epem LC 
at about 500~GeV collision energy.  
Precision measurements of its \aa width and its CP property and the 
search for its high mass partners are the main targets of the PLC.  
If a Higgs boson is not found or found at a significantly higher mass, 
I think that a PLC will be most powerful when combined with the 
highest-energy \epem LC.  
Studies of the processes $\aa \to ZZ, W^+W^-, \ttbar$ in the 
$J=0$ channel at highest \aa collision energies with high laser 
frequencies, and precise measurements of $W$ and top-quark properties 
at low laser frequencies may be most fruitful in discovering new 
physics in such cases.  

\section*{Acknowledgments}
I wish to thank Eri Asakawa for valuable discussions and for preparing 
of the figures in this report.  
I learned a lot from Eri Asakawa, Seongyoul Choi and Jae-Sik Lee through 
collaborative research works.  
Discussions with Ilya Ginzburg, Thoru Takahashi, Valery Telnov,  
Isamu Watanabe and Kaoru Yokoya have been essential for me to 
appreciate the high physics potential of a PLC.


\begin{thebibliography}{999}
\bibitem{gkst83} 
I.Ginzburg, G.Kotkin, V.Serbo, V.Telnov, {\em Nucl. Instr. Meth.\/} 
{\bf 205} (1983) 47. 

\bibitem{gkst84} 
I.Ginzburg, G.Kotkin, V.Serbo, V.Telnov, {\em Nucl. Instr. Meth.\/} 
{\bf 219} (1984) 5. 

\bibitem{telnov90} 
V.Telnov, {\em Nucl. Instr. Meth.\/} {\bf A294} (1990) 72. 

\bibitem{landauyang} 
L.F.Landau, {\em Dok. Akad. Nauk USSR} {\bf 60} (1948) 207; 
C.N.Yang, {\em Phys. Rev.\/} {\bf 77} (1950) 242. 

\bibitem{watanabe97} 
I.Watanabe et al., ``$\gamma\gamma$ collider as an option of JLC'', 
KEK Report 97-17 (March 1998). 

\bibitem{watanabe99} 
I.Watanabe, talk at LCWS99, Sitges, Spain, 28 April--5 May, 1999. 

\bibitem{jikia99} 
G.Jikia S.S\"{o}ldner-Rembold, 
hep-ph/9910366, {\em Nucl. Phys. B(Proc. Suppl.)} {\bf 82} (2000) 373.    

\bibitem{melles99} 
M.Melles, 
hep-ph/9906467, {\em Nucl. Phys. B(Proc. Suppl.)} {\bf 82} (2000) 379.    

\bibitem{chois95}
S.Y.Choi, {\em Z. Phys.\/} {\bf C68} (1995) 163; 
E.Gabrielli, V.A.Ilyin, B.Mele, 
hep-ph/9902362, {\em Phys. Rev.\/} {\bf D60} (1999) 113005;  
hep-ph/9912321; 
I.F.Ginzburg, 
hep-ph/9907549, {\em Nucl. Phys. B(Proc. Suppl.)} {\bf 82} (2000) 367; 
I.F.Ginzburg, M.Krawczyk, P.Osland, 
hep-ph/9909455, talk at LCWS99, Sitges, Spain, 28 April--5 May, 1999; 
I.F.Ginzburg, I.P.Ivanov, 
hep-ph/0004069. 

\bibitem{ginzburg00}
I.F.Ginzburg, G.L.Kotkin, 
hep-ph/9905462, {\em Eur. Phys. J.\/} {\bf C13} (2000) 295.   

\bibitem{cain}
K.Yokoya, CAIN, 
http://www-acc-theory.kek.jp/members/cain/default.html

\bibitem{lepewwg00}
A.Grutu, talk at ICHEP2000, Osaka, Japan, 27 July--2 August, 2000.  

\bibitem{susymh} 
J.R.Espinosa, M.Quir\'{o}s, 
{\em Phys. Rev. Lett.\/} {\bf 81} (1998) 516.

\bibitem{higgsatlc} 
See e.g.\ S.Yamashita, 
talk at LCWS99, Sitges, Spain, 28 April--5 May, 1999.  

\bibitem{hs93} 
K.Hagiwara, M.L.Stong, {\em Z. Phys. C.\/} {\bf 62} (1994) 99.  

\bibitem{kksz94} 
M.Kramer, J.K\"{u}hn, M.L.Stong, P.M.Zerwas, 
{\em Z. Phys. C.\/} {\bf 64} (1994) 21; 

\bibitem{hikk00} 
K.Hagiwara, S.Ishihara, J.Kamoshita, B.A.Kniehl, 
hep-ph/0002043, {\em Eur. Phys. J.\/} {\bf C14} (2000) 457.

\bibitem{gunion94}
J.F.Gunion, J.G.Kelly, 
hep-ph/9404343, {\em Phys. Lett.\/} {\bf B333} (1994) 110; 
B.Grzadkowski, J.F.Gunion, 
hep-ph/9501339, {\em Phys. Lett.\/} {\bf B350} (1995) 218;
G.J.Gounaris, G.P.Tsirigoti, 
hep-ph/9703446, {\em Phys. Rev.\/} {\bf D56} (1997) 3030.

\bibitem{choilee99} 
S.Y.Choi, J.S.Lee, 
hep-ph/9912330, {\em Phys. Rev.\/} {\bf D62} (2000) 036005. 

\bibitem{aksw99}
E.Asakawa, J.Kamoshita, A.Sugamoto, I.Watanabe, 
hep-ph/9912373, {\em Eur. Phys. J.\/} {\bf C14} (2000) 335.

\bibitem{mssmunif} 
C.Giunti, C.W.Kim, U.W.Lee, 
{\em Mod. Phys. Lett.\/} {\bf A6} (1991) 1745; 
U.Amaldi, W.de Boer, H.F\"{u}rstenau, 
{\em Phys. Lett.\/} {\bf B260} (1991) 447;
P.Langacker, M.Luo, 
{\em Phys. Rev.\/} {\bf D44} (1991) 817;
J.Ellis, S.Kelley, D.V.Nanopoulos, 
{\em Phys. Lett.\/} {\bf B260} (1991) 131.  

\bibitem{serbo00} 
V.G.Serbo, talk at $\aa$2000, DESY, Germany, 14--17 Jun 2000.  

\bibitem{achl00} 
E.Asakawa, S.Y.Choi, K.Hagiwara, J.S.Lee, 
hep-ph/0005313, to be published in {\em Phys. Rev.\/} {\bf D}.

\bibitem{hkmn90}
K.Hagiwara, K.Kato, A.D.Martin, C.K.Ng, 
{\em Nucl. Phys.\/} {\bf B344} (1990) 1. 

\bibitem{tsukamoto95}
T.Tsukamoto, K.Fujii, H.Murayama, M.Yamaguchi, Y.Okada, 
{\em Phys. Rev.\/} {\bf D51} (1995) 3153;
J.L.Feng, M.E.Peskin, H.Murayama, X.Tata, 
{\em Phys. Rev.\/} {\bf D52} (1995) 1418.

\bibitem{choizerwas} 
S.Y.Choi, A.Djouadi, H.Dreiner, J.Kalinowski, P.M.Zerwas, 
hep-ph/9806279, {\em Eur. Phys. J.\/} {\bf C7} (1999) 123; 
S.Y.Choi, A.Djouadi, H.S.Song, P.M.Zerwas, 
hep-ph/9812236, {\em Eur. Phys. J.\/} {\bf C8} (1999) 669; 
S.Y.Choi, M.Guchait, J.Kalinowski, P.M.Zerwas, 
hep-ph/0001175, {\em Phys. Lett.\/} {\bf B479} (2000) 235;
S.Y.Choi, A.Djouadi, M.Guchait, J.Kalinowski, H.S.Song, P.M.Zerwas, 
hep-ph/0002033, {\em Eur. Phys. J.\/} {\bf C14} (2000) 535.

\bibitem{feng96}
N.Arkani-Hamed, H.C.Cheng, J.L.Feng, L.J.Hall, 
hep-ph/9603431, {\em Phys. Rev. Lett.\/} {\bf 77} (1996) 1937; 
J.J.Cao, T.Han, X.Zhang, G.R.Lu, 
hep-ph/9808466, {\em Phys. Rev.\/} {\bf D59} (1999) 095001.

\bibitem{feng97}
H.-C.Cheng, J.L.Feng, N.Polonsky, 
hep-ph/9706438, {\rm Phys. Rev.\/} {\bf D56} (1997) 6875;
hep-ph/9706476, {\rm Phys. Rev.\/} {\bf D57} (1998) 152; 
J.L.Feng, 
hep-ph/0002055, {\em Int. J. Mod. Phys.\/} {\bf A15} (2000) 2355.

\bibitem{hikasa95}
K.Hikasa, Y.Nakamura, 
hep-ph/9501382, {\em Z. Phys.\/} {\bf C70} (1996) 139; 
Erratum {\bf C71} (1996) 356.  
 
\bibitem{nojirim96}
M.M.Nojiri, K.Fujii, T.Tsukamoto, 
hep-ph/9606370, {\em Phys. Rev.\/} {\bf D54} (1996) 6756; 
M.M.Nojiri, D.M.Pierce, Y.Yamada, 
hep-ph/9707244, {\em Phys. Rev.\/} {\bf D57} (1998) 1539; 
S.Kiyoura, M.M.Nojiri, D.M.Pierce, Y.Yamada, 
hep-ph/9803210, {\em Phys. Rev.\/} {\bf D58} (1998) 075002. 

\bibitem{ch95} 
S.Y.Choi, K.Hagiwara, 
hep-ph/9506430, {\em Phys. Lett.\/} {\bf B359} (1995) 369; 
S.Y.Choi, K.Hagiwara, M.S.Baek, 
hep-ph/9605334, {\em Phys. Rev.\/} {\bf D54} (1996) 6703.

\bibitem{jikiazz} 
G.V.Jikia, 
{\em Phys. Lett.\/} {\bf B298} (1993) 224;
{\em Nucl. Phys.\/} {\bf B405} (1993) 24.

\bibitem{jikiaww} 
G.Jikia, 
hep-ph/9612380, {\em Nucl. Phys.\/} {\bf B494} (1997) 19.

\bibitem{jikiatt}
B.Kamal, Z.Merebashvili, A.P.Contogouris, 
hep-ph/9503489, {\em Phys. Rev.\/} {\bf D51} (1995) 4808; 
G.Jikia, A.Tkabladze, 
hep-ph/9601384, {\em Phys. Rev.\/} {\bf D54} (1996) 2030;
G.Jikia, A.Tkabladze, 
hep-ph/0004068.

\bibitem{gounaris99} 
G.Jikia, A.Tkabladze, 
hep-ph/9312228, {\em Phys. Lett.\/} {\bf B323} (1994) 453;
hep-ph/9312274, {\em Phys. Lett.\/} {\bf B332} (1994) 441;
G.J.Gounaris, P.I.Porfyriadis, F.M.Renard, 
hep-ph/9812378, {\em Phys. Lett.\/} {\bf B452} (1999) 76; 
hep-ph/9902230, {\em Eur. Phys. J.\/} {\bf C9} (1999) 673;
G.J.Gounaris, J.Layssac, P.I.Porfyriadis, F.M.Renard, 
hep-ph/9904450, {\em Eur. Phys. J.\/} {\bf C10} (1999) 499; 
hep-ph/9909243, {\em Eur. Phys. J.\/} {\bf C13} (2000) 79. 

\bibitem{watanabe92}
Y.Yasui, I.Watanabe, J.Kodaira, I.Endo, 
hep-ph/9212312, {\em Nucl. Instrum. Meth.\/} {\bf A335} (1993) 385. 

\bibitem{telnov93}
V.Telnov, Proc.\ of LCWS93, Waikola, Hawaii, 26--30 April 1993 
(World Scientific, 1993) p.551.  

\end{thebibliography}
\end{document}